# Small-Size Resonant Photoacoustic Cell of Inclined Geometry for Gas Detection

**A.V.Gorelik[a], A.L.Ulasevich[a], F.N.Nikonovich[a], M.P.Zakharich[a], A.I.Chebotar[b], V.A.Firago[c], V.M.Stetsik[c], N.S.Kazak[a], V.S.Starovoitov[a,d]**

[a]B.I.Stepanov Institute of Physics, NASB, Nezavisimosti Ave. 68, 220072 Minsk, Belarus
[b]KBTEM-OMO, Planar corporation, Partizansky Ave. 2, 220763 Minsk, Belarus
[c]Belarus State University, Nezavisimosti Ave. 4, 220030 Minsk, Belarus
[d]e-mail: vladstar@dragon.bas-net.by

**Abstract.** A photoacoustic cell intended for laser detection of trace gases is represented. The cell is adapted so as to enhance the gas-detection performance and, simultaneously, to reduce the cell size. The cell design provides an efficient cancellation of the window background (a parasite response due to absorption of laser beam in the cell windows) and acoustic isolation from the environment for an acoustic resonance of the cell. The useful photoacoustic response from a detected gas, window background and noise are analyzed in demonstration experiments as functions of the modulation frequency for a prototype cell with the internal volume $\sim 0.5$ cm$^3$. The minimal detectable absorption for the prototype is estimated to be $\sim 1.2 \cdot 10^{-8}$ cm$^{-1}$ W Hz$^{-1/2}$.

**PACS** 07.07.Df; 42.62.Fi; 82.80.Kq

## 1. Introduction

Laser-based photoacoustic spectroscopy is one of the most sensitive techniques applied to local non-contact analysis of trace amounts of various chemical compounds in gas media [1-3]. The most appropriate light sources for this technique are lasers operated in the infrared wavelength region. They are powerful carbon dioxide (the oscillation wavelength region 9 - 11 µm) and carbon monoxide (2.8 - 4.1 or 5.1 - 8.0 µm) CW lasers [4,5] , CW parametric oscillators (3 - 5 µm) [6], quantum cascade lasers (4 – 5 µm) [7,8] and miniature DFB laser diodes (0.7 - 2 µm) [9]. The photoacoustic technique is distinguished by a high sensitivity at the (sub) ppb (parts per billion, $1{:}10^9$) level with the minimal detectable absorption of $10^{-10}$ cm$^{-1}$, a short time resolution (down to a few seconds) and the capability to recognize reliably a large number of chemical compounds [2,3]. The infrared-laser-based photoacoustic technique is generally recognized to be a reasonable approach to the inexpensive *in situ* multi-component analysis of atmospheric air in environmental pollution monitoring [2,10,11], exhaust gas monitoring [2,12,13], industrial process control [14,15] and leak detection [16]. The technique finds an expanding application in biology-related areas (plant physiology [3], crop and fruit storage [17], entomology [18] and medical diagnostics by means of exhalation analysis [19,20]) for sensitive detection of gases being involved in the metabolic reactions and emitted by living systems.

The miniaturized spectroscopic hardware of trace-gas analysis is a promising field of application for the photoacoustic technique. The principle of the technique is based on measuring the amplitude and phase for the acoustic pressure oscillation (the so-called photoacoustic response) arising due to absorption of a modulated laser beam by molecules of gas inside a photoacoustic cell. The photoacoustic detection is realized with an enhanced sensitivity if the modulation frequency coincides with an acoustic resonance of the cell. Therefore, the modulation frequency should be correlated with the cell-resonator sizes: the frequency increases with reducing the sizes. Implementation of laser-beam modulation at ultrasonic frequencies allows to reduce the cell sizes down to a few millimeters. Combining the high sensitivity inherent in the traditional photoacoustic approaches with an ability to probe the gas inside such a small-sized photoacoustic cell gives a possibility to analyze chemical compounds to be emitted by individual small-sized objects with an extremely low emission rate. A crude estimation shows that the application of the approach can provide detecting the gas leak emitted by an object with the rate down to $\sim 10^{-14}$ cm$^3$/s for $C_2H_4$ or $\sim 10^{-10}$ cm$^3$/s for $CO_2$. For comparison, in the aerobic reaction an individual cell of living organism emits $CO_2$ with a rate from $10^{-10}$ to $10^{-9}$ cm$^3$/s [21]. In the photosynthesis reaction an individual cell of plant can absorb $CO_2$ at a rate higher than $10^{-8}$ cm$^3$/s [22]. The detection sensitivity will be increased $10^3 - 10^6$ times compared to the traditional photoacoustic approaches and more than 100 times in relation to the commercial mass-spectrometer-based leak-detection systems. In contrast to the systems, the ultrasonic photoacoustic leak detector needs no expensive vacuum chambers and can be applied to *in situ* localization of leak for a large number of substances to be emitted in atmospheric air. According to the theoretical estimations (see, for instance, [23-25]), the amplitude of photoacoustic response can be increased with reducing the cell sizes.

We accept that the miniaturized photoacoustic cell must not be worse in the performance in comparison to non-miniaturized one. The cell performance is specified usually in terms of the signal-to-noise ratio. In order to obtain a high magnitude for the ratio the cell design must satisfy some requirements. First of all, the design has to provide detection of the highest possible useful signal (that is, a photoacoustic response from the gas inside the cell) at a selected acoustic resonance. The cell must be reliably isolated from external acoustic noise and parasite electric pickups. And, finally, the background effect of cell windows on the measurements should be reduced to a minimum.

The background effect requires a special attention for the miniaturized cells. This effect is associated usually with absorption of laser beam in the cell windows resulting to local heating of the gas inside the cell. The measurement inaccuracy arisen from this effect is proportional both to the laser power instability and to the power absorbed by the windows. If a non-miniaturized photoacoustic cell is used the effect can be critical for high-power laser applications (for instance, at an intracavity arrangement of the cell to a carbon dioxide or carbon monoxide laser operated with the intracavity power of some tens Watts). The background effect is not very significant for non-miniaturized low-power laser systems (such as laser diodes or quantum cascade lasers operated at a power up to a few tens of milliWatts). But, our experience testifies that the size reduction for the photoacoustic cell can result in a dramatic increase in the background signal. This increase can be observed for a low-power laser beam. For instance, the window background signal is observed in our experiments for a small-sized photoacoustic cell at a laser beam power down to 10 mW. This sentence does not contradict the theoretical conclusions [23-25].

The quartz-enhanced photoacoustic spectroscopy (QEPAS) is a reasonable approach to miniaturize the spectroscopic hardware with the help of the increased-frequency modulation of laser beam [26]. Instead of a gas-filled resonant acoustic cavity, the sound energy is accumulated in a high-Q quartz crystal frequency standard. Usually the standard is a quartz tuning fork with an acoustic resonant frequency of 32 kHz in air. To the moment, a great success is achieved in QEPAS-based gas detection with the help of compact cells

(the internal cell volume reaches down to a few cubic centimeters) and different near- or mid-infrared laser systems (including diode and quantum cascade lasers, optical parametric oscillators) [27-29]. The experiments testify that the high-sensitivity photoacoustic detection can be realized at ultrasonic modulation frequencies if the rates of intra- and intermolecular collisional vibration-vibration (VV) or vibration-translation (VT) energy redistribution for the detected species are high compared to the modulation frequency.

The goal of the work is to reveal the application potential for another possible approach to the miniaturized resonant photoacoustic cell. In contrast to the QEPAS technique, we support a traditional approach to photoacoustic detection: through a small hole in the cell shell, a condenser microphone registers the photoacoustic response to the laser beam modulated with the frequency of an acoustic resonance of the cells. According to the approach, an efficient way to miniaturize the resonant photoacoustic cell can be implemented with the help of a properly adapted design of the cell. Here we represent a simplified cell design and describe some helpful principles, which allow to optimize the design for the best gas-detection performance. We demonstrate also how such an approach can be applied in practice. An experimental examination of a compact prototype cell is performed. In the examination we estimate the measurement error associated with the noise in the absence of laser beam and test the acoustic isolation of the cell from the environment. Then we analyze the photoacoustic signals (they are the useful photoacoustic signal and window background) to be initiated by a laser beam inside the cell. The performance of the cell is estimated in terms of the minimal detectable absorption.

## 2. Cell design

Here we describe a resonant photoacoustic cell of simple design. The cell is intended for detection of gases with the help of a collimated and linearly polarized laser beam. Geometrically, the cell is similar to a ring, the face and end planes of which are cut to the optical axis at an angle $\vartheta$. The optical windows are attached to the face and end planes. The cell design is represented in Figure 1 as a section of the cell by a plane formed by the optical axis OO' and the electric polarization vector **P** for the laser beam. The internal cavity of the cell is specified in terms of the diameter $d$ of clear aperture and thickness $h$ of the ring. In order to cancel the parasite reflection of laser beam from the cell windows the angle $\vartheta$ is accepted to be equal to the Brewster angle. Due to the inclined geometry, the acoustic eigen-modes for the cell cavity can not be classified as longitudinal or transverse modes (for instance, as radial or azimuthal ones for the cylindrical resonator [23]). We identify the cell eigen-modes $\nu_n$ by the number of nodes $n$ given by intersection between the relevant standing wave and the longest cell diagonal (the diagonal AB in the Figure 1). The mode $\nu_1$ (the mode with the lowest non-zero eigen-frequency $\omega_1$) is specified by one node on the diagonal. The mode $\nu_2$ (the eigen-frequency $\omega_2$) gives two nodes on the diagonal. This mode is specified by 3 anti-nodes located at acute corners of the cell cavity (near the ends A and B of the line AB) and near the cell midpoint (the point given by intersection of the axis OO' with the line AB).

The parameters (they are the diameter $d$ and thickness $h$), which specify the geometry shape for the cell cavity, the microphone location and arrangement of inlet/outlet gas holes are fitted in order to enhance the cell performance for an individual acoustic eigen-mode. The parameters are fitted with the help of a numerical simulation. In the simulation the spatial amplitude distribution for an acoustic standing wave, which answers to the selected mode, is determined. From this distribution we find the useful signal (a photoacoustic response from the gas inside the cell) and window background (a signal caused by absorption of the laser beam in the cell windows). We estimate also the effect of external acoustic noise on the measurements. The simulated signals are analyzed as functions of the geometry parameters. The parameters are fitted in such a way as to increase the useful signal and, simultaneously, to minimize the negative role of window background and external noise for the selected mode. In a sense, we 'manipulate' by the node-antinode structure of spatial mode distribution in order to enhance the cell performance. The analysis of the simulated data testifies that the most efficient parameter optimization can be realized for the acoustic mode $\nu_2$. Therefore, the cell design is optimized for the acoustic mode $\nu_2$.

In the optimization we follow some simple rules. First of all, we fit the ratio between the diameter $d$ and thickness $h$. A proper choice of the ratio leads to the mutual elimination of photoacoustic signals generated from the front and end windows and, as a result, to the total cancellation of window background. Secondly, according to the numerical simulation, the maximal amplitude for the standing wave of mode $\nu_2$ is realized at antinodes located near the acute corners of the cell cavity (see Figure 1). The simulated wave amplitude in the corners is nearly twice as high as that in the cell midpoint. Therefore, the microphone is arranged near an acute corner of the cell. Such a position for the microphone allows also to analyze the acoustic properties for others eigen-modes. Third, the inlet and outlet gas holes provide a gas flow through the cell. They are located near nodes of the mode $\nu_2$. Such an arrangement of the holes is intended to reduce the negative influence of the holes on the acoustic Q-factor of the mode $\nu_2$ and to isolate the measurement from external acoustic noise at modulation frequencies close to $\omega_2$.

We have applied the represented optimization procedure to make a miniaturized version for the cell. The cell windows are made of polycrystalline ZnSe. The internal cell volume is ~ 0.5 cm$^3$ and the diameter $d$ of clear aperture is 8 mm. The length of optical path inside the cell is no longer than 1 cm. A miniature condenser Knowles FG-3629 microphone (sensitivity 22 mV/Pa at 10 kHz, noise ~ 54 nV/Hz$^{1/2}$ at 14 kHz) is applied for detection of the acoustic signals. The microphone is connected to the internal cell cavity by a duct (the duct diameter of 0.7 mm, the duct height of 1 mm). The diameter for inlet and outlet holes in the cell wall is 0.3 mm. Our numerical simulation predicts that in the frequency range from 0 to 20 kHz we can observe three acoustic resonances (near $\omega_1 = 8.1$ kHz, $\omega_2 = 14.2$ kHz and $\omega_3 = 19.2$ kHz) associated with eigen-modes $\nu_1$, $\nu_2$ and $\nu_3$. This prototype cell is applied in experiments where we analyze the acoustic properties of the cell and estimate the cell performance.

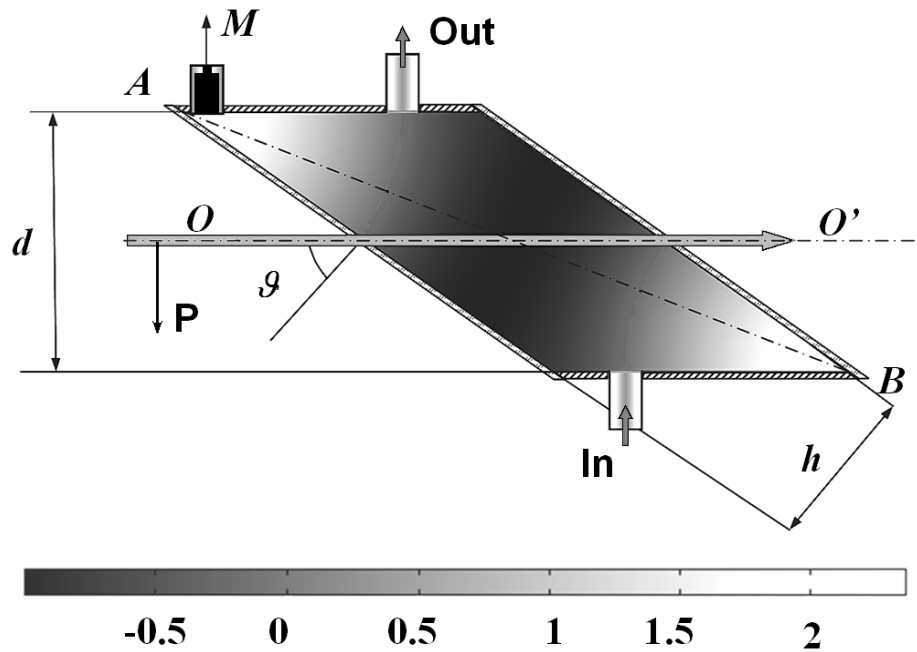

Figure 1. The cell design: OO' - optical axis, **P** - electric polarization vector for the laser beam, M – microphone, In – gas inlet port, Out - gas outlet port. The wave pattern (grey filling) inside the cell cavity demonstrates a spatial amplitude distribution for the acoustic standing wave of mode $\nu_2$.

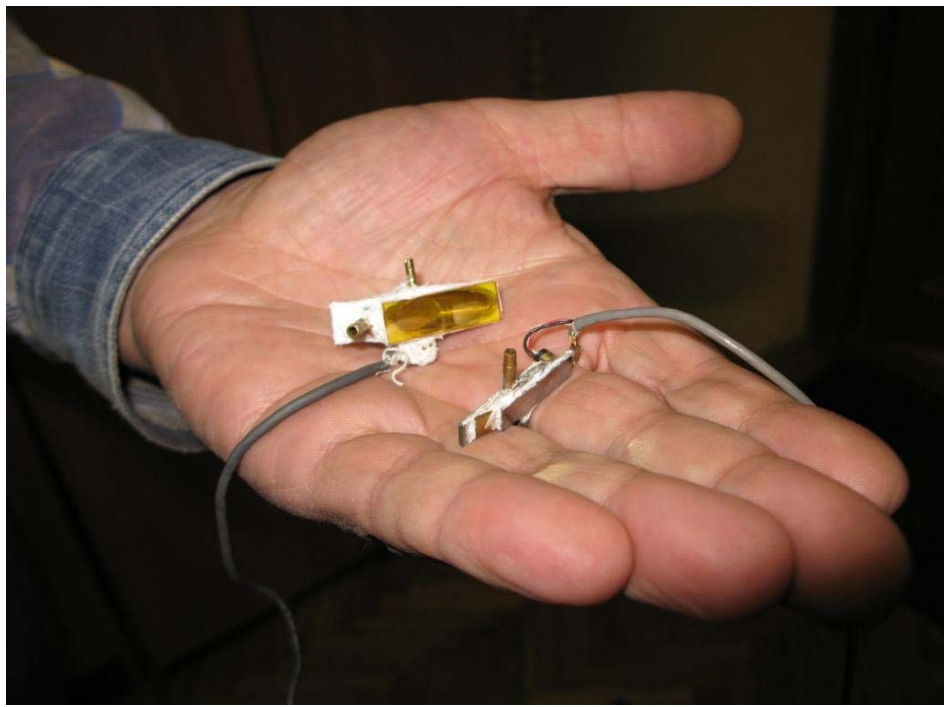

Figure 2. Photoacoustic cells of inclined geometry with Brewster windows. The left cell (*ZnSe* windows) has the internal volume V ~ 500 mm$^3$ and clear aperture $d = 8$ mm. This cell is applied in the report as a prototype cell. The right one (*Ge* windows) is a cell with V ~ 80 mm$^3$ and $d = 4$ mm.

## 3. Noise analysis

We estimate the measurement error associated with the acoustic or electric noise registered by the microphone inside the prototype photoacoustic cell in the absence of laser beam. We assume that the cell cavity can be coupled acoustically with the environment through inlet and outlet gas ducts. Therefore, in the estimation we take into account a possible manifestation of acoustic disturbances produced outside the cell by a source.

The cell is inserted into a capsule, which provides a reliable sound isolation from the noise produced in our laboratory room (see Figure 3). The capsule design allows the connection of the cell cavity with the environment through only the inlet and outlet flexible gas ducts of the inner diameter > 1 mm. The compressed air blown through a nozzle is used as a source of external acoustic disturbances. These disturbances imitate the 'white' noise for the considered frequency range. The nozzle is placed near the suction hole of inlet gas duct. The conditioned laboratory air is applied as a gas medium filling the cell cavity.

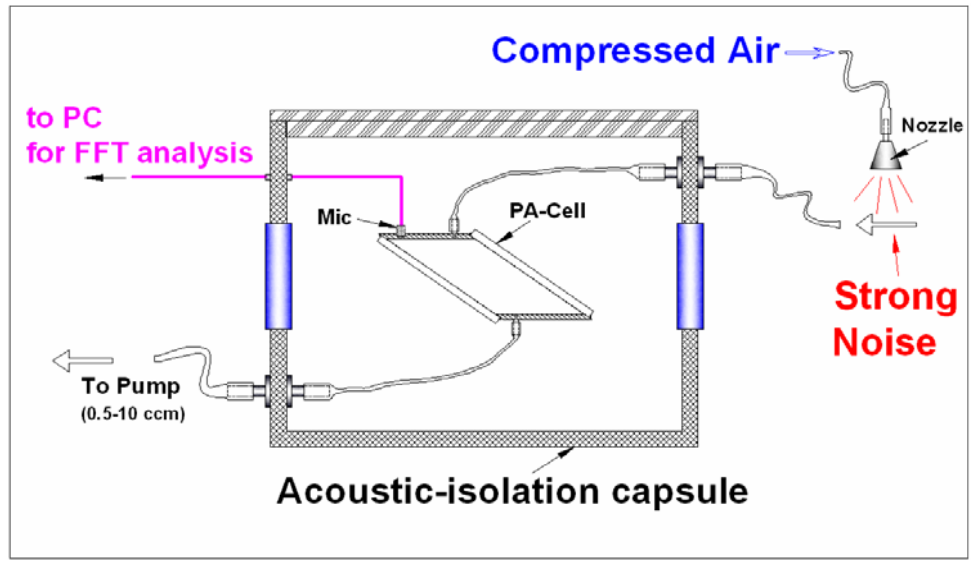

Figure 3. The experimental arrangement for estimation of the measurement error associated with noise (Mic - microphone)

The estimation of measurement error is performed for acoustic signals detected at different strengths of the external-noise effect. The negligibly small influence of external noise on the measurements is realized when the source of external disturbances is switched off. The inlet and outlet gas ducts are closed and the cell cavity is acoustically isolated from the environment. A regime of strict silence is kept in the laboratory room. The strong effect of external noise on the measurements is produced if the source of external disturbances operates. The inlet and outlet gas ducts are opened and allow the gas flow through the cell at flow rates higher than 30 cm$^3$/s (a flow rate, which provides a perfect gas renewal inside the cell for the time ~ 1 sec).

In the experiment we estimate the measurement error in terms of a measure of inaccuracy $\sigma_\omega$ for the Fourier transform spectrum $S_\omega$ of detected acoustic signals. The value of error $\sigma_\omega$ is determined as a standard root-mean-square deviation for the Fourier transform $S_\omega$ of acoustic signal:

$$\sigma_\omega = (<S_\omega S_\omega^*> - <S_\omega><S_\omega^*>)^{1/2}$$

The quantity $S_\omega$ is calculated with the help of the fast-Fourier-transform procedure performed for individual time-sample realizations of acoustic signal (each realization represents the microphone signal over a given time interval $\tau_{avr} \approx 0.13$ s). The symbol $<...>$ means the averaging over the ensemble of the realizations. In the experiment the number of the signal-sample realizations is more than 1000.

We analyze the deviation $\sigma_\omega$ as a function of frequency $\omega$ for the frequency range from 0 to 22 kHz. Figure 4 demonstrates two typical frequency-dependences for $\sigma_\omega$ obtained at different levels of external acoustic noise. When the external noise is negligible small the obtained dependence for $\sigma_\omega$ corresponds to the minimal error attainable by the acoustic signal. The deviation $\sigma_\omega$ is a slowly varying function of $\omega$. Slight resonant dependences are observed when the frequency is close to eigen-frequencies of acoustic modes $\nu_1$, $\nu_2$ and $\nu_3$. The magnitude of $\sigma_\omega$ is in the interval of values from $10^{-7}$ to $4 \cdot 10^{-7}$ V. We associate such a behavior of $\sigma_\omega$ with the electric noise of microphone and fluctuations of gas pressure inside the cell.

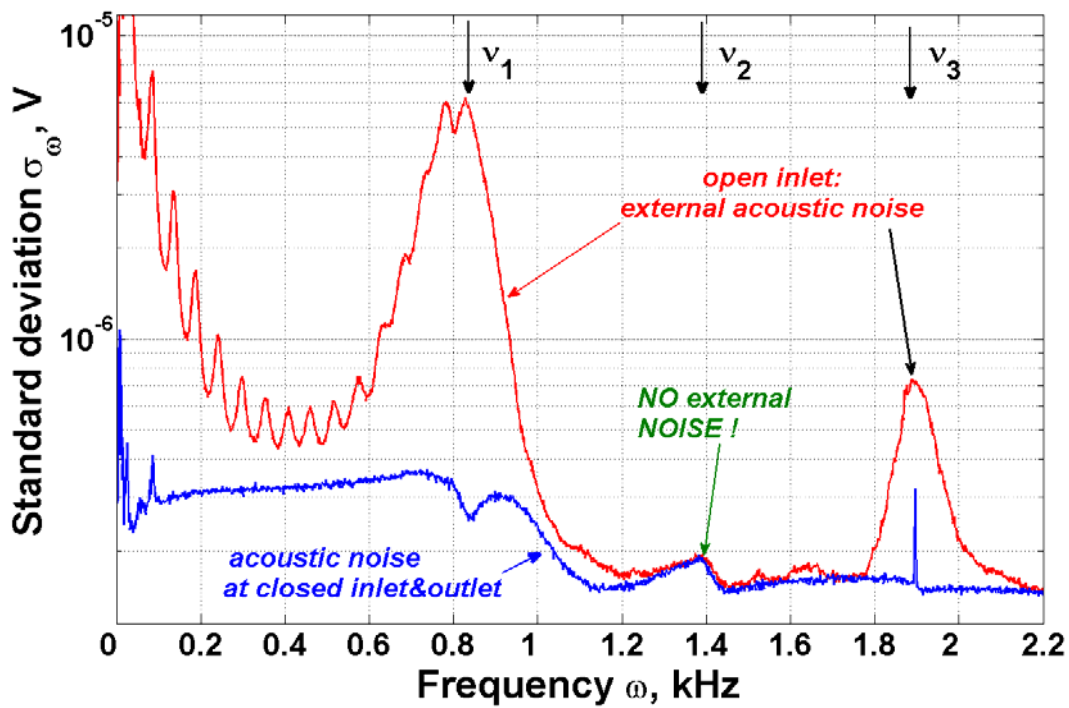

Figure 4. The frequency dependence of the standard deviation $\sigma_\omega$ for the Fourier transform of acoustic signals detected in the absence of laser beam at two different levels of external acoustic disturbances. The solid blue line gives the deviation $\sigma_\omega$ obtained when the effect of external acoustic noise is negligible small (the inlet and outlet gas ducts are closed). The solid red line gives the deviation $\sigma_\omega$ associated with the measurement at strong external acoustic noise (the inlet and outlet gas ducts are opened). The arrows point the location of resonant frequencies for the acoustic modes $\nu_1$

In the presence of strong external noise the frequency dependence of $\sigma_\omega$ is essentially transformed. Frequency regions, which answer to acoustic resonances of the cell, are susceptible strongly to the external disturbances. For these regions, the effect of external disturbances can result in a considerable increase for the measurement error. The disturbances lead to a substantial increase in $\sigma_\omega$ for low-lying frequencies (the frequencies located near zero eigen-frequency [23] within the interval from 0 to 2 kHz), which correspond to the application region for the so-called non-resonant photoacoustic detection. The observed oscillation in $\omega$-dependence of $\sigma_\omega$ at low frequencies is associated with manifestation of acoustic resonant properties for the flexible inlet and outlet gas ducts. A considerable growth of $\sigma_\omega$ with the disturbances is exhibited also for the frequencies located near the resonances with the modes $\nu_1$ and $\nu_3$. But, no increase of $\sigma_\omega$ is observed for the frequencies resonant to the mode $\nu_2$. Our experiment testifies that for the frequencies $\omega \approx \omega_2$ the standard deviation $\sigma_\omega$ is not affected by the external disturbances. Regardless the strengths of applied disturbances, the quantity $\sigma_\omega$ does not exceed the value $2 \cdot 10^{-7}$ Volts. We associate this mode immunity to the disturbances with a proper arrangement of inlet and outlet gas holes drilled in the cell wall. Notice that such an acoustic isolation for an individual acoustic mode can be implemented when the external disturbances are applied to the cell through only the inlet and outlet gas ducts.

## 4. Analysis of useful and background signals

In order to analyze the laser-initiated photoacoustic signals we perform an experiment where a collimated beam of tunable carbon dioxide laser is applied. A detailed description of the used experimental set-up is given in [30,31]. In the experiment we detect the useful and background signals initiated by the laser beam on individual oscillation lines of carbon dioxide laser. The useful signal is accepted to be a photoacoustic response generated due to absorption of laser beam by trace amounts of ammonia in nitrogen. The laser *'on-line'* wavelength to be used for ammonia detection corresponds to the vibration-rotation line 9R(30) of band $00^01$-$[10^00,02^00]_{II}$ of $^{12}C^{16}O_2$ molecule. According to the Hitran database (http://www.cfa.harvard.edu/hitran/), the absorption $\alpha_{on}$ on this line for ammonia is 73.9 cm$^{-1}$atm$^{-1}$. Such a strong absorption implies that the photoacoustic signal detected on the line can be accepted reliably as a useful response from the gas to be analyzed provided that the ammonia content in gas mixture is higher than 1 ppm. All measurements in the experiment are made for flows of ultra-high purity nitrogen (99.9995 %), which contains ammonia at concentrations from 5 to 15 ppm. The background signals are detected with the help of measurements on the laser line 9R(24) (*'off-line'*) of the same band. According to the Hitran database, the absorption on this line for ammonia is negligible small (0.001 cm$^{-1}$ atm$^{-1}$) in order to have any influence on the detected signal for the applied ammonia concentration. The rate of gas flow to be blown through the cell is maintained automatically with the help of a flow controller in such a way to provide the perfect gas renewal inside the cell for the time $\sim 1$ sec. The pressure and temperature of gas in the cell is accepted to be close to typical parameters for the laboratory room (740 torr and 22 degrees Celsius). The aperture of the laser beam inside the photoacoustic cell is $\sim 0.5$ mm. The beam is modulated with the help of a Germanium acousto-optic modulator. In the absence of ammonia in the gas flow the photoacoustic *'off-'* and *'on-line'* signals are identical.

In the experiment we analyze the *'on-'* and *'off-line'* signal amplitude and phase as functions of the modulation frequency $\omega$. The signals are detected in the presence of ammonia in the gas flow for the frequency range from 1 to 20 kHz. Figure 5 shows obtained frequency dependences for the amplitude in Volts. The *'on-'* and *'off-line'* signals exhibit the different behavior on the frequency scale. Admixing ammonia to the flow at *'on-line'* measurements leads to an increase in the signal amplitude. The *'on-line'* signal demonstrates clearly resonant peaks when the modulation frequency is close to $\omega_1$, $\omega_2$ or $\omega_3$ predicted by the numerical simulation. The analysis of the obtained data testifies that the Q-factor $Q_2 \approx 12.3$ ($Q_2 = \omega_2/\Delta\omega_2$, $\Delta\omega_2$ is the peak width at half maximum) for the resonant peak in the vicinity of $\omega_2$ is considerably higher compared to the relevant parameter $Q_1 \approx 4.6$ for the acoustic resonance near $\omega_1$. We explain such an improved performance for the acoustic mode $\nu_2$ by a proper location of the inlet and outlet gas holes in the cell wall.

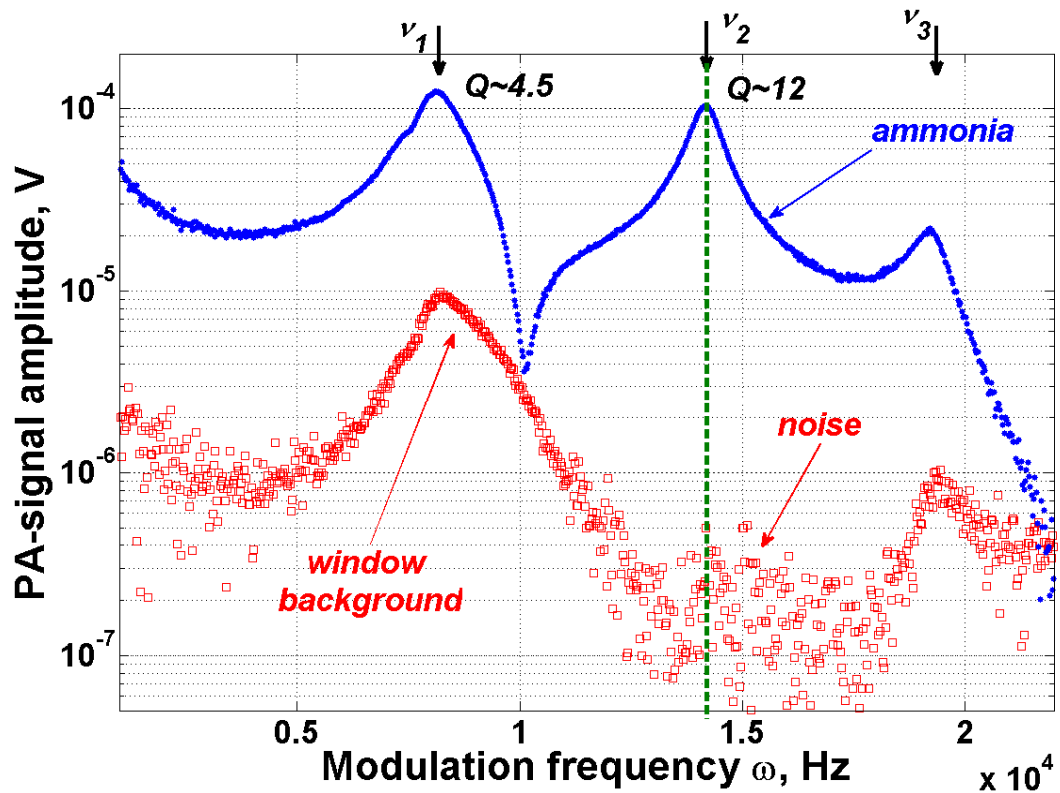

Figure 5. The amplitude of photoacoustic signal obtained from '*on-line*' (blue full circles) and '*off-line*' (red open squares) measurements as a function of the modulation frequency $\omega$. The measurements are performed on individual oscillation lines of carbon dioxide laser for nitrogen flow admixed by ammonia (13.6 ppm) [30,31]. The '*on-line*' power (oscillation laser line 9R(30)) is 66 mW. The '*off-line*' power (line 9R(24)) is 130 mW. The time of averaging for every point is 0.13 s. The arrows point the location of resonant frequencies for the acoustic modes $\nu_1$, $\nu_2$ and $\nu_3$.

Whatever the ammonia concentration in the gas flow, the frequency dependence of the '*off-line*' signal amplitude remains invariable. Contrary to the '*on-line*' dependence, the '*off-line*' signal manifests only two resonant peaks. They are acoustic resonances near the frequencies $\omega_1$ and $\omega_3$. We associate the manifestation of these peaks with a background signal due to absorption of laser beam in the cell windows. No resonance peak in the vicinity of $\omega_2$ is observed. Moreover, at $\omega \approx \omega_2$ the '*off-line*' signal is strongly noised and the signal amplitude attains a minimum, which is comparable with the standard deviation $\sigma_\omega$ (the root-mean-square error) obtained for the detected acoustic signals in the absence of laser beam. The detailed analysis of the '*off-line*' signal measured at the modulation frequency $\omega \approx \omega_2$ shows that the signal phase is distributed in a quasi-uniform random manner over the range from 0 to $2\pi$. It implies that at $\omega \approx \omega_2$ the magnitude of the window background is negligible small in comparison to the noise.

Our study testifies that the optimization of the parameters, which specify the design of internal cavity for the photoacoustic cells of inclined geometry, leads both to an improvement of resonant properties and to a minimization of the window background for the acoustic mode $\nu_2$. The minimal error of measurements is attained at a modulation frequency $\omega \approx \omega_2$ and when the measurement error is due to the acoustic and electric noises of microphone.

## 5. Estimation of cell performance

Obviously, the gas-detection sensitivity for the cell is maximal when the modulation frequency $\omega$ is close to $\omega_2$. Therefore we estimate the cell performance at this frequency. The performance is specified in terms of the normalized noise equivalent absorption (NNEA). We accept this quantity as the minimal detectable absorption $\alpha_{min}$, which can be obtained if the signal-to-noise ratio is equal to 1 provided that the laser power is 1 W and the averaging time is 1 sec. The coefficient $\alpha_{min}$ is found from:

$$\alpha_{min} = \alpha_{on} \, C_{NH3} \, \tau_{avr}^{1/2} \, P_{on} \, N_{off}/S_{on}$$

Here the quantities $C_{NH3}$ and $\tau_{avr}$ denote actual values for ammonia concentration and the averaging time. The power $P_{on}$ is an '*on-line*' power for the modulated laser beam before the front cell window. The parameter $S_{on}$ gives the amplitude for the '*on-line*' photoacoustic signal at $\omega = \omega_2$. The quantity $N_{off}$ corresponds to the root-mean-square error for the acoustic signal obtained from the '*off-line*' measurements. According to our measurements, the signal-to-noise ratio $S_{on}/N_{off}$ is 1095 ($C_{NH3}$ = 12.36 ppm, $\tau_{avr}$ = 0.13 s, $P_{on}$ = 49 mW). The minimal detectable absorption $\alpha_{min}$ is estimated to be $\sim 1.2 \cdot 10^{-8}$ cm$^{-1}$ W Hz$^{-1/2}$.

## 6. Conclusion

Thus, we have presented a resonant photoacoustic cell of simple design. The design geometry for the cell is specified by a few parameters, which can be optimized in order, for instance, to adapt the cell to experiment needs and to enhance the cell performance. Certainly, the design simplicity makes the parameter optimization and cell manufacturing easy and extends the area of cell application in practice. The cell sizes can be reduced, hereafter, down to a few millimeters by a trivial scaling procedure. The obtained results testify to a great potential for the miniaturized photoacoustic cells of the inclined geometry in creating compact ('pocket-sized') high-sensitivity laser sensors of chemical compounds. Such a millimeter-sized cell powered, for instance, by a minute DFB laser diode with a typical output power of $\sim$ 20 mW will allow to detect trace gases at a sensitivity level, which can be attained by the absorption-probing spectroscopy techniques along optical paths of some kilometers. The miniature background-free and 'acoustically isolated' photoacoustic cell with Brewster windows ca be successfully inserted into the cavity of a high-power laser in order to find a fascination application for *in-situ* detection of the smallest gas leaks. To the moment, the performance of our miniaturized cells is not high in comparison to the relevant parameter for non-miniaturized ones. Nevertheless, we expect to enhance the performance in the nearest future. We go on [32].

## Acknowledgments

This work was performed in the framework of B-1252 project supported by International Science and Technology Center.